\documentclass[12pt,ams]{iopart}
\usepackage{cite}
\usepackage{latexsym}
\usepackage{graphicx}
\usepackage{rotating}
\usepackage{longtable}
\usepackage[usenames]{color}
\usepackage[normalem]{ulem}

\input colordvi

\begin{document}

\title{Self-organization in Pd/W(110): 
interplay between surface structure and stress}

\author{N.~Stoji\'{c}$^{1,2}$, T.~O.~Mente\c{s}$^3$ and N.~Binggeli$^{1,2}$}

\address{$^1$Abdus Salam International Centre for Theoretical Physics, 
Strada Costiera 11, Trieste 34151, Italy}
\address{$^2$ IOM-CNR Democritos,  Trieste, I-34151, Italy}
\address{$^3$ Sincrotrone Trieste S.C.p.A., Basovizza-Trieste
  34149, Italy}
\ead{nstojic@ictp.it}

\date{\today}

\begin{abstract}

It has recently been shown that submonolayer Pd on W(110) forms
highly-ordered linear mesoscopic stripes at high
temperatures. The stripes
display an internal Pd superstructure with a nano-scale periodicity
along the direction perpendicular to the periodicity of the stripes.
The same type of superstructure is also observed in a wide range of
temperatures below the stripe formation temperature. We present a
combined experimental and theoretical study of this superstructure of
Pd on W(110) and investigate its influence on the appearance of the
linear mesoscopic stripes. By means of low-energy electron diffraction
and low-energy-electron microscopy we show that it has a far more
peculiar dependence on temperature and coverage than expected from a
regular surface reconstruction. Using density-functional theory, we
model the Pd superstructures  as periodic  vacancy-line type of
configurations and investigate  their energetics and elastic
properties.  From our calculated surface stresses and
anisotropies for the vacancy-line type of configurations, and based on the
continuum elasticity theory, we demonstrate quantitatively that the
vacancy-line type of  structure is a prerequisite for the formation of
the linear mesoscopic stripes. Moreover, we show that the physics
driving the formation of the internal superstructure is very similar
to the one at play in forming the mesoscopic stripes themselves.

\end{abstract}


\maketitle

\section{Introduction}

Recently, a novel mesoscopic striped phase of Pd/W(110) was observed
above 1000~K ~\cite{MenLocAba08}, for submonolayer Pd
coverages. 
It is long known that spontaneously formed periodic patterns can be
stabilized at the mesoscopic scale via an energy balance due to long-ranged  strain
fields and short-ranged atomic bonds at the interface of two
subdomains, \cite{AleVanMea88} as has been demonstrated in many
examples so
far~\cite{FigLeoBar08,MenStoLoc11,GasPlaBar03,KerNieSch91,EllRepRou01,MenLocAba08}.
In general, the inherent stress at 
surfaces of crystalline materials is an important driving factor 
in deciding the structural properties at length scales spanning from 
atomic to macroscopic
distances~\cite{Ibachs,MulSau04,SanTiaKir09,StoBin12}. A recent
investigation of the two-component Pd-O adsorbate stripes on
W(110)~\cite{MenStoLoc11}
demonstrated how stripes of various shape and orientation can be
obtained by altering the oxygen concentration and the resulting 
surface stress differences between the two adsorbate
phases.

Interestingly, the Pd stripes in the Pd-O-covered W(110) display also themselves 
an internal  nano-scale  
superstructure~\cite{MenStoLoc11}.
This structure is of the same type as displayed by submonolayer
Pd on W(110)~\cite{MenStoLoc11,SchBau80}, which has been previously described as
one-dimensional pseudomorphism  at room temperature~\cite{SchBau80}. 
Also for the Pd/W(110) stripes  the same type of internal
structure can be deduced, based on the similarity of the low-energy
electron diffraction (LEED) patterns of the Pd/W(110) mesoscopic
stripes~\cite{MenLocAba08} and the submonolayer Pd on W(110) 
superstructure~\cite{SchBau80}. Based on these two independent
findings~\cite{MenLocAba08,SchBau80}, it can be  inferred that
the periodicities of  the mesoscopic linear stripes and their internal
superstructure are along two perpendicular directions, which is a
peculiar feature. Moreover, it can be expected 
that the internal superstructure has an important, if not decisive, role in
the linear mesoscopic stripe creation, as the existence of such stripes is conditioned
by a limited range of surface stress differences between the two phases
on the surface~\cite{GaoLuSuo02}.

Overall, there have been only a few structural studies of ultrathin 
 Pd films on W(110)~\cite{MenLocAba08,SchBau80,SanPueCer10,RifShiKim09,AbaBarLoc04}. For
 submonolayer to monolayer Pd coverages, this includes a combined LEED and
 Auger electron spectroscopy study~\cite{SchBau80}, a core-level
 photoemission study~\cite{RifShiKim09} and spectromicroscopic studies
 of the mesoscopic structures based on low-energy electron microscopy
 (LEEM)~\cite{MenLocAba08,AbaBarLoc04} and x-ray photoelectron
 microscopy (XPEEM)~\cite{AbaBarLoc04}. In their
extensive work,  Schlenk and
Bauer~\cite{SchBau80} reported a periodic superstructure along
$[1\bar{1}0]$ for  submonolayer  Pd on W(110) annealed at  temperatures above 450~K. The 
superstructure LEED spots observed in this
study remained independent of coverage up to about 0.9~monolayer (ML), above which
it  transformed into a pseudomorphic (PS) layer. 
However, further details on the nature of this one-dimensional
pseudomorphism of submonolayer Pd on W(110) are still missing.
The questions in this regard become yet more interesting considering
the lack of similar superstructures in 
submonolayer Pd films on related bcc(110) surfaces of Ta and
Nb~\cite{BolCar91}.

In this paper, we present  a combined experimental and theoretical
study of the Pd/W(110) periodic  superstructure, and investigate theoretically
the influence of this structure on the formation of the linear
mesoscopic stripes. 
We first study the structure of Pd on W(110) as a function of coverage
and temperature using LEED and LEEM. Our LEED measurements reveal, in
particular, a striking continuous evolution of the Pd superstructure
period, as a function of temperature, with an extrapolated
inverse-period trend leading to the PS structure at 0~K. In order to
understand the presence and nature of the superstructure, we then
investigate, by means of {\it ab~initio} density-functional theory (DFT)
calculations, the equilibrium atomic structure and energetics of
selected adlayer Pd configurations on W(110) compatible with the LEED
characterization. The calculations consider the PS structure along
with    periodic vacancy-line type of superstructures. Among all
configurations theoretically considered, the PS one is singled out as
the lowest-energy state at 0~K, although the vacancy-line
  structures lie very close in energy. The differences are within the
  thermal energy for the temperatures at which the vacancy-line
  structures are observed. The experimental observation of the
formation of the Pd superstructure, instead of the PS layer,
at high temperature is explained qualitatively to be due to vacancy
formation and vacancy-vacancy interactions becoming more pronounced at
elevated temperature. Finally, by means of DFT we also examine the
influence of the vacancy lines on the Pd/W(110) surface stress and
demonstrate that the presence of the vacancy lines of
nano-scale periodicity is a prerequisite for the formation of the
observed Pd linear mesoscopic stripes. We note that
our experimental findings on Pd/W(110) are similar to the observations on submonolayer
Au/W(110)~\cite{Geo_thesis},
which point to the generality of the forces at play in the formation of the superstructure,
as we will
discuss in the following.

The paper is organized as follows: in Section II we describe the
experimental and theoretical methods. In Section III,  we present our
LEED results on Pd/W(110) as a function of Pd coverage and
temperature. Section IV containts our results for the atomic structure
and energetics of the PS and vacancy-line  superstructures, as well
as a discussion on the origins of the latter type of structures. While
the results in section IV are mainly based on DFT calculations, we
also report there LEEM and LEED data for the PS Pd/W(110) structure
confirming the theoretical Pd adsorption site. Section V, based entirely
on theory, includes our
DFT results on the influence of the vacancy lines on the Pd/W(110)
surface stress, and a discussion of the resulting dependence of the
mesoscopic stripe pattern on the vacancy-line period of the 
superstructure. The Conclusions are then given in Section VI.

\section{Methods}

\subsection{Experiment}

The LEED and LEEM
measurements were performed using the SPELEEM (Spectroscopic
PhotoEmission and Low Energy Electron Microscope) at the
Nanospectroscopy beamline (Elettra, Italy)~\cite{LocAbaMen06}.  LEEM
operates by imaging elastically-backscattered electrons. Along with 
the usual contrast mechanisms~\cite{Bau98}, the ability to filter
electrons at the diffraction plane allows
to apply dark-field methods to LEEM. The spatial resolution in imaging
is 12~nm.

In the micro-diffraction mode, the instrument is capable of acquiring
LEED patterns from a 2~$\mu$m-sized region. The transfer  width of the
microscope is about 10~nm, which defines the angular resolution.  The
distortion of the diffraction pattern is corrected by an offline
analysis, allowing a precise determination of the distances in k-space.

The W(110) substrate was cleaned by the standard procedure of exposing
it to $1\times10^{-6}$~mbar of molecular oxygen at 1100$^\circ$ C, and
subsequent flashes to high temperature to remove oxygen. Pd was
deposited from a target heated by electron bombardment. The rate
calibration was done by following the changes in electron reflectivity.

\subsection{Theory}

We have performed DFT pseudopotential  calculations in a plane-wave
basis, using the PWscf code, a part of the Quantum
espresso distribution~\cite{GiaBarBon09_short}.   The  Perdew-Zunger
parametrization~\cite{PerZun81} of the  local-density approximation
(LDA) has been adopted for exchange and correlation.
To simulate different surface adlayer configurations, we used a
supercell approach. An asymmetric slab with 1~Pd layer on
5~layers of W substrate (bottom 2~W layers were fixed) and
9~vacuum layers was constructed, both for relaxations and for subsequent
stress calculations. The only exception related to the substrate thickness are
the calculations of the pseudomorphic 
structure, which were performed on a 15-layer asymmetric slab, with
2 bottom W layers fixed.

Vanderbilt ultra-soft pseudopotentials~\cite{Van90} were generated from
the $4d^9 5s^1 5p^0$ atomic configuration of  Pd and  from $5s^2 5p^6
5d^4 6s^2$  configuration of W. The core-cutoff radii  for Pd were: 
 $r_{s,p}=2.4$~a.u. and $r_d=1.8$~a.u., and for W: $r_{s,p}=2.2$, $r_d
 = 2.4$~a.u. Our kinetic energy cutoff was 35~Ry for the wave
 functions and 350~Ry for the charge density. A $34\times34\times1$ k-point  
 Monkhorst-Pack mesh~\cite{MonPac76} centered at $\Gamma$ was used for the
 $1\times 1$ W(110) surface unit cell. The analogues of that mesh with constant
 k-point density were used for the supercells of different dimensions. 
  The theoretical W lattice
 constant, $a= 3.14$~\AA~ was used  in the calculations (the
 experimental value is 3.16~\AA). Total energy
 differences were converged to better than 0.5~meV per 6-layer slab
 (1 Pd + 5 W layers) with $1\times1$ W surface unit cell lateral
 extent, while the forces were
 converged to better than 3.5~mRy/\AA. 

The surface stress was computed using the analytical expression 
derived by Nielsen and Martin~\cite{NieMar83}, based on the 
Hellmann-Feynman theorem. The surface stress uncertainty was estimated
to be 0.36~N/m on the basis of convergence tests  
regarding the kinetic-energy and wave-function-cutoffs, the 
number of k-points and the number of vacuum layers~\cite{MenStoBin08}.

In order to obtain surface stress values for Pd-terminated surface
from our assymetric slabs (with Pd on top and frozen W on the bottom),
we subtracted the reference surface stress of the frozen W surface
from the stress obtained for the assymetric slab. 
We checked the accuracy of this procedure by evaluation of surface
stress from a symmetric slab (11 layers) and an assymetric (6 layers) slab
calculation. 
The differences were within the estimated 
uncertainty, giving a confirmation of the validity of our calculational
approach. We also checked the convergence of surface stress with
respect to the slab thickness, as 
it is shown in Supplementary data.  By increasing the number of vaccum layers,
we also checked that the presence of an electric field, due to different terminations
of our slab, has no significant influence on the surface stress results. 

For comparison, we also performed stress calculations for the Pd(111)
surface. The theoretical lattice constant of bulk Pd is
$a_{Pd}=3.88~$\AA~ (experimentally, it is 3.89~\AA). In fact, the PS
configuration of the Pd overlayer on W(110) can be viewed as a Pd(111)
monolayer oriented with its $[10\bar{1}]$ inplane axis parallel to the
W[001] surface axis, and expanded by 14.4\% along the W[001]
direction and contracted by 6.6\% along the W$[1\bar{1}0]$ axis, with
respect to the bulk Pd(111) layer (the corresponding experimental
values are $\epsilon^{[001]}=+14.9$\% and
$\epsilon^{[1\bar{1}0]}=-6.2$\%). For the Pd(111) surface calculations, we
used a symmetric slab of 11 Pd layers (with the 4 outer layers on each
side relaxed) and the same plane-wave cutoffs and k-point grid as for
the W surface. 


\section{LEED  Results on Pd/W(110)}

\begin{figure}[ht]
\begin{center}
\includegraphics[width=7cm]{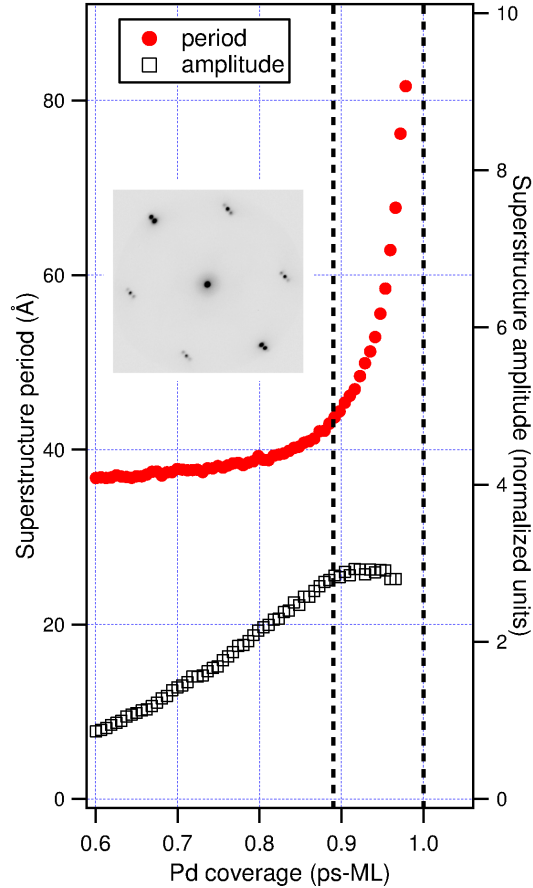}
\caption{  Pd superstructure period, and the amplitude
of the superstructure spot (normalized to the integral
beam intensity) as a function of Pd coverage at 750 K. 
Inset shows the LEED pattern with the extra Pd spots. The vertical
dashed lines mark the  completion of the loosely packed mismatch layer
and that of the denser pseudomorphic layer in sequence.}  
\label{fig:exp}
\end{center}
\end{figure}

We monitored the evolution of the Pd mismatch to the underlying 
W lattice, via the extra Pd diffraction spots separated along $[1\bar{1}0]$
from those of the substrate  (see inset in figure~\ref{fig:exp}).
For all temperatures considered, the spot profiles were well described
by gaussians used in finding the peak positions. Moreover, the sharp
spots confirm the presence of a long-range ordered superstructure.
Figure~\ref{fig:exp}  displays the evolution of the superstructure
period as a function of Pd coverage,  along with the corresponding
changes in the intensity ratio between the superstructure and substrate spots 
acquired at a fixed temperature of 750~K.
The real-space superstructure period is obtained by taking the
inverse of the diffraction spot separation.

\begin{figure}[t]
\begin{center}
\includegraphics[width=10cm]{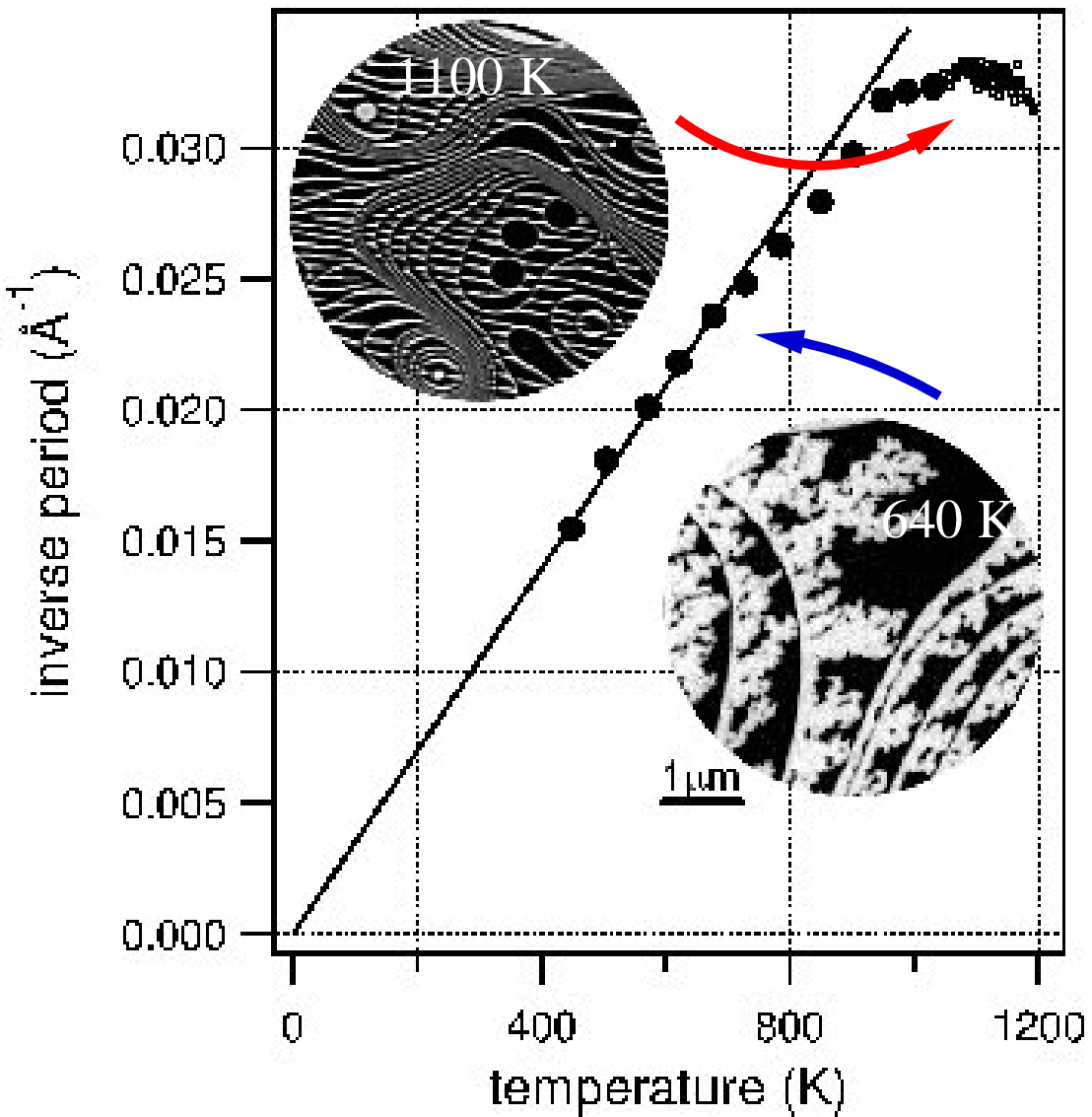}
\caption{ Inverse period, determined from LEED, of the Pd superstructure as a function of
  temperature at 0.5~ML Pd coverage. The solid guideline passing through the data points
  highlights the trend towards the pseudomorphic structure ({\em i.e.}, zero
  mismatch) at 0~K. Also shown are two LEEM images of the same surface taken at
  different temperatures. The linear mesoscopic stripes are clearly
  seen in the image taken at 1100~K.  }  
\label{fig:exp_temp}
\end{center}
\end{figure}

The coverage dependence of the period 
shows two important characteristics. Firstly, the period is almost the same for all 
coverages below about 0.89~ML at this temperature, consistent with the
past measurements~\cite{SchBau80}. 
The Pd extra spots are separated from those of the underlying tungsten by a 
constant momentum transfer for all diffraction orders. 

The second feature is the continuous transition of the Pd lattice towards the 
pseudomorphic structure above 0.89~ML. The
coverage at which the transition starts  
is marked by the first vertical dashed line in figure~\ref{fig:exp}. 
Attributing this coverage ($0.89 \pm 0.04$~ML\cite{note_transition})
to the superstructure layer covering the whole surface, 
we can deduce that there are two Pd vacancies per superstructure
period (about 45~\AA). Note that, 
at different temperatures, this transition, although not a sharp one,
systematically shifts in coverage according to the  
dependence of period on temperature, further supporting the assignment
of two vacancies per period. 
Also importantly, the transition to the pseudomorphic structure as a
function of coverage takes place without an  
accompanying broadening of the diffraction peaks. This would indicate a continuous 
increase in the period of the defect line structure due to the additional material. 

The distinction of Pd  superstructure from a regular reconstruction
is underlined by its  temperature dependence. We observe that the
inverse period continuously decreases for decreasing temperatures,
going from 0.032 to 0.015~\AA$^{-1}$ as the temperature changes from
900~K to 400~K as seen in figure~\ref{fig:exp_temp}. The tendency of the
inverse period towards zero with decreasing temperature is consistent
with the pseudomorphic structure as the limiting case at 0~K. The
superstructure spot  shows a nearly constant full-width at
half maximum  in the range 400 to 900~K\, as
shown in the Supplementary Information. Furthermore, below
400~K, the width increases sharply
and the Pd spot cannot be identified reliably.  
We attribute this to a kinetic limitation, as the period gets longer
whereas the diffusion is hindered at lower temperatures. 
Above 900~K, the spot again broadens considerably, and the structure period
remains nearly constant. This roughly coincides with the experimental onset of
stress-induced mesoscopic stripe formation~\cite{MenLocAba08}, as also
monitored here by LEEM (see insets in figure~\ref{fig:exp_temp}).
A similar change in the behaviour of spot profiles at about 900~K is observed in Au/W(110) and
is attributed to a lattice solid-liquid transition~\cite{Geo_thesis}.

\section{Pd-adlayer atomic structure and energetics}
\subsection{ Pseudomorphic structure: theoretical and experimental results}

\begin{figure}[t]
\begin{center}
\includegraphics[width=9cm]{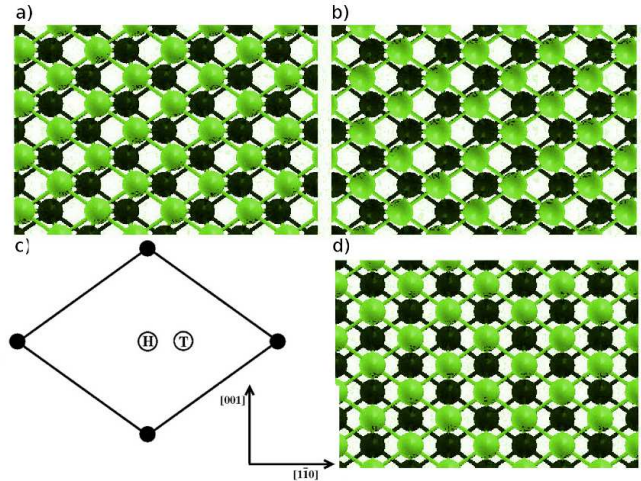}
\caption{   Top view of the ground-state pseudomorphic
  structure with (a) R and (b) L   adsorption site. The T and H adsorption sites are shown in
  (c), while the H-site PS structure is shown in (d). Only the upper
  two layers (Pd+W) are displayed;  Pd and W atoms are
shown in green (grey) and black, respectively. }
\label{fig:PS_struct}
\end{center}
\end{figure}

We have investigated the atomic structure of the pseudomorphic
Pd/W(110) layer using DFT calculations. Figures~\ref{fig:PS_struct}(a) and
(b) describe the  atomically-relaxed configurations.  Although
previous density-functional calculations generally assumed the centro-symmetric
``hollow'' (H) adsorption site for Pd on (110) bcc metal
surfaces~\cite{WuFre95,WuCheKio96}, we find that 
the preferred adsorption site is very close to the three-fold coordinated
(T) adsorption site, shown in
figure~\ref{fig:PS_struct}(c). The actual adsorption site is displaced
by 0.5~\AA~ from the H site in the direction of the T site, and is only
 0.05~\AA~ away from the T site,  in
figure~\ref{fig:PS_struct}(c). Breaking of the $(1\bar{1}0)$ mirror symmetry 
signifies  the existence of two domains degenerate in energy,
those with left (L) and right (R) adsorption sites, relative to the H
site, as shown in figure~\ref{fig:PS_struct}(a) and (b). The lateral
displacements,  with respect to the atomic positions in the
continuation of the ideal W bcc lattice, as well as
the interlayer distances of the adsorbate layer and of the three upper W
layers  are given in Table~\ref{table:PS_displacements}. The 
atoms of the upper W layer (and of the two layers below) are
only slightly displaced from the bulk positions as can be seen in
Table~\ref{table:PS_displacements}. The energy difference between the
H-site configuration (figure~\ref{fig:PS_struct}(d)) and ground state
for the Pd PS layer (figures~\ref{fig:PS_struct}(a) and (b))  is 67
meV per Pd atom~\cite{note_d_H_site}.

\renewcommand{\baselinestretch}{1}
\begin{table}[ht]
\bigskip
\begin{center}
\caption{Atomic displacements along   [1$\bar{1}$0] (x) relative to the
  bulk positions in the continuation of the W bcc lattice and changes in
interlayer distances, relative to the bulk interlayer distance,
for the upper 4 layers of the 
PS configuration with a 3-fold adsorption site (ground state).}
\begin{tabular}{ c c c}
\hline
\hline \\[-2mm]
Layer & $\Delta x$ (\AA) 
       & $\Delta d_{n,n+1}$ (\%) \\ [2mm]
\hline  \\[-3mm]
 $Pd$       & $0.50$          & $-6.1$     \\
 $W^{1st}$   & $-0.02$         &  $-0.9$     \\
 $W^{2nd}$   & $0.00$          &  $-0.2$     \\
 $W^{3rd}$   & $0.01$          &  $-0.1$     \\
\hline
\hline
\end{tabular}
\label{table:PS_displacements}
\end{center}
\end{table}

\begin{figure}[t]
\begin{center}
\includegraphics[width=9cm]{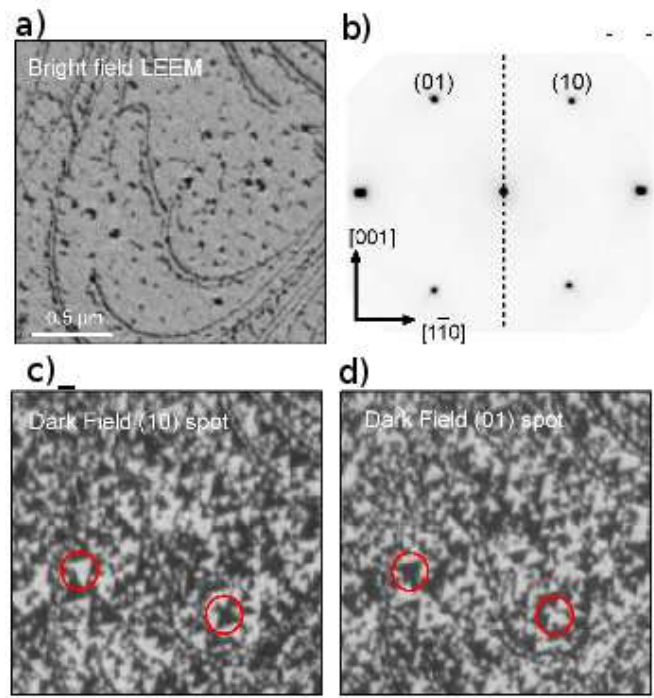}
\caption{ (a) LEEM image of the PS Pd monolayer on
  W(110). The curved lines are tungsten single atomic steps, and the
  dark spots are 3D islands on top of the Pd monolayer. (b) Corresponding LEED pattern. 
The diffraction spots coincide with those of the bcc(110) tungsten substrate. 
(c) and (d) Dark field LEEM images acquired by imaging with the (10)
and (01) LEED spots as marked in (b). The presence of triangular
domains indicates the broken mirror symmetry about the [001] axis
marked by the dashed vertical line on the LEED pattern.}
\label{fig:PS_struct_exp}
\end{center}
\end{figure}

The breaking of the $(1{\bar 1}0)$ mirror symmetry  is directly confirmed
by LEEM measurements displayed in figure~\ref{fig:PS_struct_exp}. The LEEM image
obtained from the specularly-reflected electrons ({\em i.e.}, bright-field
LEEM) shows, in 
figure~\ref{fig:PS_struct_exp}(a), the pseudomorphic
Pd monolayer along with small islands of larger thickness. 
The coverage was chosen on purpose to be slightly above 1~PS-ML,
in order to avoid the complicated  superstructures below monolayer coverage.

The LEED pattern in figure~\ref{fig:PS_struct_exp}(b) confirms that the Pd
layer has the same in-plane symmetry and period as the W(110) surface.
However, the dark field images in figures~\ref{fig:PS_struct_exp}(c) and (d)
reveal  the presence of two domains. The dark-field images are
acquired by using a filter at the diffraction plane in order to image
with only those electrons belonging to a particular diffraction spot.
The two panels displayed in figure~\ref{fig:PS_struct_exp} are acquired
using the integral spots connected by a mirror reflection about the
[$001$] axis.  The presence of complementary domains reflects the absence of
the corresponding mirror symmetry.

\subsection{ Vacancy-line structure: DFT results}

\begin{figure}[th]
\begin{center}
\includegraphics[width=6.1cm]{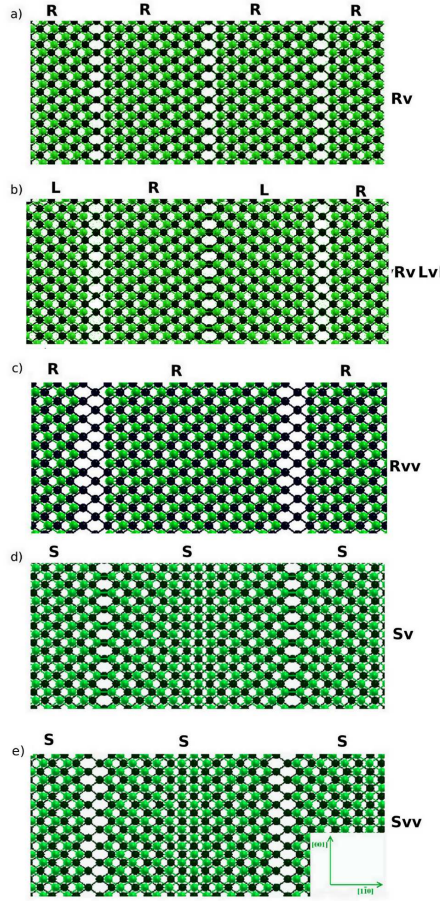}
\caption{ Examples of vacancy-line superstructures considered in our
  DFT study, with vacancy lines along [001].
 The upper two layers (Pd+W) of the  relaxed atomic
  structures are shown. The structures are labeled according to the
  sequence, along [1$\bar{1}$0], of Pd (R, L, S) domains and vacancy
  lines (v, vv), as indicated on the right-hand side of the figure
  (see also text). The
  period and coverage of the superstructures are indicated using the
  notation $n:m$, where $n$ is the supercell period (number of W atoms)
  along [1$\bar{1}$0] and $m$ is the number of Pd atoms per period:
  (a) 10:9 Rv,  (b) 20:18 RvLv, (c) 16:14 Rvv, (d) 16:15 Sv, and (e) 16:14 Svv configurations.   
The atom colors and directions are like in figure~\ref{fig:PS_struct}.   }
\label{fig:mismatch_struct}
\end{center}
\end{figure}
For coverages below 1~ML, the superstructure observed in LEED indicates a smaller
density of Pd atoms along [1$\bar{1}$0] than in the pseudomorphic structure.
In our DFT simulations, atomic relaxations systematically indicate the spontaneous 
creation of a vacancy line  when the number of Pd atoms is less than
the number of registry positions. This is found for various initial configurations,
including  starting from a homogenous distribution of Pd atoms, with
interatomic Pd distances along [1$\bar{1}$0] larger than the W-W distance
in that direction. Therefore, we have modeled the
observed superstructure with periodic vacancy lines pointing along the
[001] direction. We note that we also 
 compared the  formation energy of vacancy lines along [001]
and along [1$\bar{1}$0], using a small test structure with a period of
6 W surface unit cells {\em i.e.}, 5 Pd atoms per 6 adsorption sites (assuming one
adsorption site per W surface unit cell). The vacancy
formation energy along [1$\bar{1}$0] is larger by 75~meV/vacancy than
along [001], which is in agreement with the experimentally observed direction for
the vacancy lines, and gives the energy scale corresponding to the
anisotropy of this superstructure.

Figure~\ref{fig:mismatch_struct} illustrates the types of vacancy-line
structures, with vacancy lines along [001], which we considered. They are
labeled according to their periodic sequence, along  [1$\bar{1}$0], of
Pd domains and vacancy lines (such as Rv, RvLv, etc.), where ``R''
(``L'') refers to a Pd domain with preferred R- (L-) three-fold sites
and ``v'' stands for the vacancy line. As indicated in
figure~\ref{fig:mismatch_struct}, we considered single (``v'') and
double (``vv'') vacancy lines, and also symmetric (``S'') Pd domains.
The period and Pd coverage of the superstructure are specified with
the notation $n:m$, where $n$ is the supercell period (number of W atoms)
along [1$\bar{1}$0] and $m$ is the number of Pd atoms per period,
yielding the superstructure nominal coverage: $\eta = m/n$. We consider up to
two vacancies per period and, for simplicity (symmetry), R and L
domains of the same size in a given superstructure. For simplicity, we
consider only straight vacancy lines, neglecting roughening effects~\cite{Per92}.

As mentioned earlier, the LEED data displayed in figure~\ref{fig:exp}
indicate two Pd vacancies per superstructure period.
Therefore, we principally focused on structures with $n:(n-2)$, 
which are made of either a double-vacancy line, 
or two single-vacancy lines that separate alternating L and R
domains. For the sake of comparison, however, we also included  in our
study single-vacancy configurations with $n/2:(n/2-1)$, for even $n$,
which correspond to the same coverage $\eta$ as the $n:(n-2)$
configurations.

\begin{figure}[th]
\begin{center}
\includegraphics[width=7cm, angle=270]{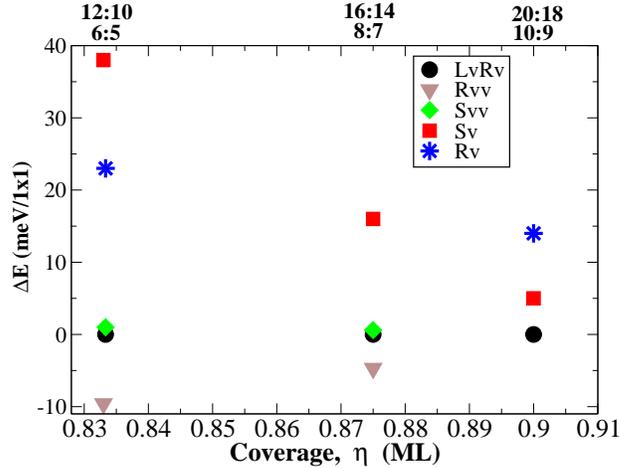}
\caption{ Relative energy of different configurations as a function of
  their Pd coverage $\eta$.
  The RvLv configuration is taken as the reference for all three
  coverages considered.
  The configurations which correspond to the three
  coverages are indicated at the top of the figure. 
}
\label{fig:size_effect}
\end{center}
\end{figure}

We considered three different coverages between 0.8 and 0.9~ML and
several different configurations for each of these coverages.  
The relative total energies of these various configurations (with
relaxed atomic structure) at each coverage are displayed in
figure~\ref{fig:size_effect}, as a function of coverage. Two of these 
coverages, namely the ones related to the 20:18 and 16:14 configurations, are
compatible with experiment, as shown in figure~\ref{fig:exp_temp}, at temperatures of
$\sim$650~K and 850~K, respectively. For the highest coverage, fewer
theoretical configurations were considered, as the relaxations for the
$n=20$ supercell size become computationally very heavy. Inspection of
figure~\ref{fig:size_effect} indicates that, consistent with the LEED
measurements (showing two vacancies per period) the structures with
one vacancy per period (Sv, Rv) are systematically highest in energy. 
We also observe that the double-vacancy structure (Rvv) is the lowest
in energy for the two lowest coverages considered. In view of the
trend, in figure~\ref{fig:size_effect},  of decreasing energy
differences (approaching zero) with  increasing coverage (to 1~ML),
one may also expect the Rvv configuration to be slightly lower in
energy than the RvLv and Svv configurations at the third coverage
(0.9~ML). However, it should be noted that
 the energy differences between the Rvv, RvLv and
Svv are very small, namely within 5~meV per Pd atom for the second
coverage in figure~\ref{fig:size_effect}, {\em i.e.}, for the period of 16 W
[1$\bar{1}$0] lattice spacings (the period observed at about
850~K, in figure~\ref{fig:exp_temp}).  
Hence, although the Rvv structure is theoretically slightly lower in
energy, all three structures are essentially degeneratre within kT, at
the high temperature (T~=~850~K) at which they are observed (and we
expect to be same for the third period corresponding to $\eta = 0.9$
and T~=~650~K)~\cite{note_degrees_freedom}.

\begin{figure}[ht]
\begin{center}
\includegraphics[width=7.5cm, angle=0]{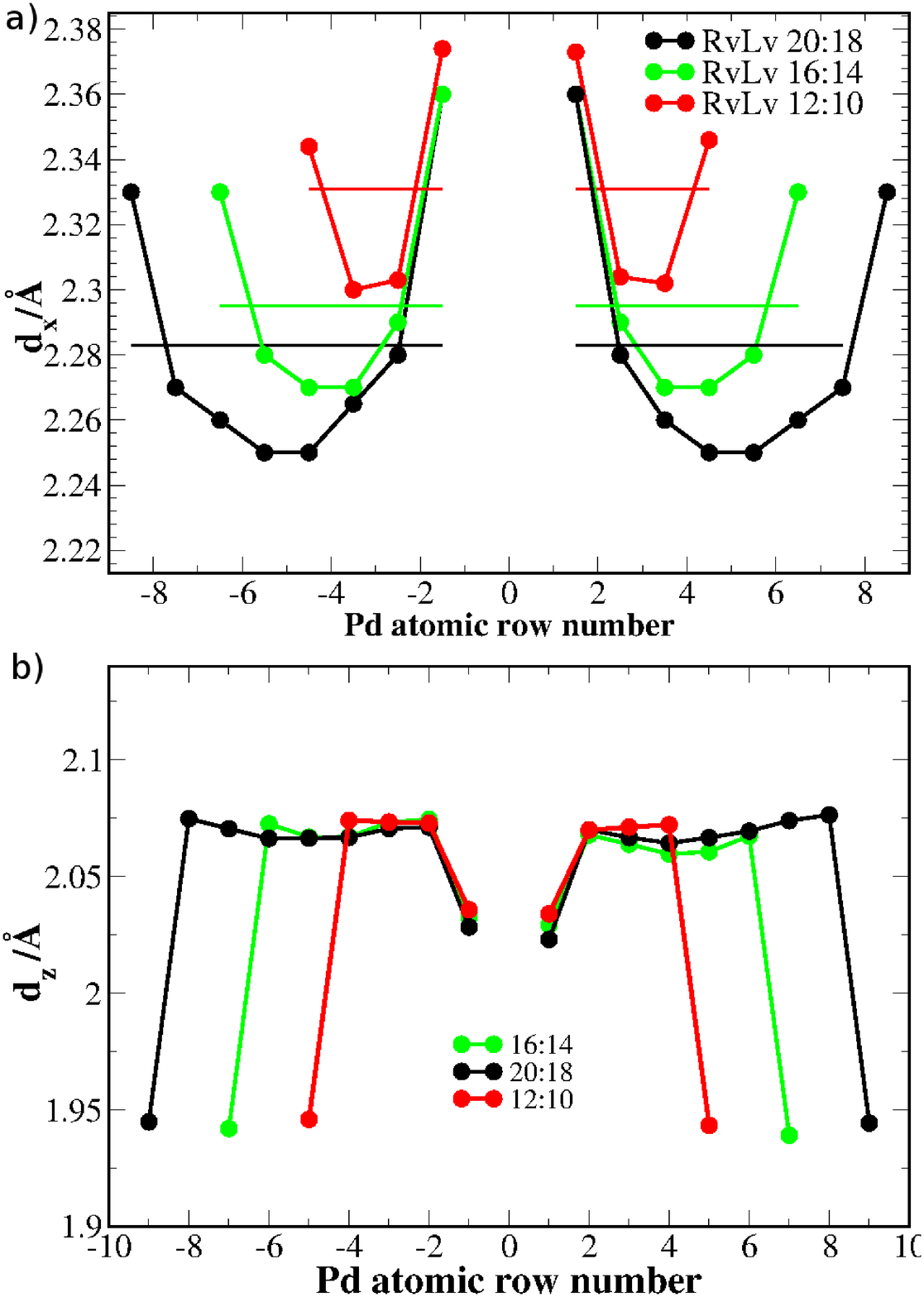}
\caption{Interatomic spacing, $d_x$, along  [1$\bar{1}$0] between rows
  of Pd atoms (a) and positions of the Pd atoms perpendicular to the
  W(110) surface (b) in the  RvLv type of configurations with
  different superstructure periods.  For comparison, 
$d_x$ for the W(110) surface is 2.22~\AA. Straight lines indicate  the
average values.
}
\label{fig:3fold_pos}
\end{center}
\end{figure}

The relaxed atomic configurations reported in figure~\ref{fig:mismatch_struct} disclose
the tendency of the Pd atoms to ``slide'' on the tungsten surface
along the  [1$\bar{1}$0] axis, leading to Pd adsorption
sites which gradually change throughout the R, L, and S structures in between two
grain boundaries. To describe the magnitude of these
variations of the Pd overlayer and their spatial distribution, we plot
in figure~\ref{fig:3fold_pos} 
the optimized Pd  interatomic spacings
in the  [1$\bar{1}$0]  direction,  $d_x$,  (a) and Pd atomic
positions, $d_z$, perpendicular to the W(110) surface (b) for RvLv structures  
with periodicities 12, 16, and 20. (The Rv, Rvv, Sv and Svv structures are 
considered in the Supplementary data).
The positions $d_z$ are measured relative to the outermost W atom within
the Pd-covered region of the surface. 
``Atom row number'' describes the atomic row position within
the surface  structure in the [1$\bar{1}$0] direction;  the zero
is located at the center of the narrow vacancy-line between the R and L adsorption
sites ({\em i.e.}, the central vacancy in
figure~\ref{fig:mismatch_struct}(b)).
  We see, from figure~\ref{fig:3fold_pos}(b), that large
relaxations in the vertical direction occur only at the edges of the
Pd domains, and
that the central part of the Pd domains 
does not have significant
variations along the $z$ direction. The main relaxation is in $d_x$
(figure~\ref{fig:3fold_pos}(a)) and is a systematic  ``stretching'' of
the Pd domains in the
[$1\bar{1}0$] direction along  
which vacancy lines are introduced. The largest expansion takes
place around the vacancy lines, 
while the least stretched are the central parts of the Pd domains. 
Figure~\ref{fig:3fold_pos}(a) also shows
that the average Pd-Pd distance (stretching) increases with decreasing
size of the Pd domains. Such a stretching of the Pd domains is a
general feature we find for all superstructures considered with [001]-vacancy
lines (see Supplementary data). The Pd(111) layer is indeed
compressed ($-6.6$~\% lattice mismatch, see Section IIB) along the
[1$\bar{1}$0] axis in
the PS configurations, and tends to relax the associated strain in the
presence of vacancy lines. Such vacancy lines may thus be viewed as
analogous to dislocation lines at fully developed interfaces. 

We also note that, in the specific case of the Rv and RvLv structures
which are rather similar (figures.~\ref{fig:mismatch_struct}(a) and (b)),
the relaxation of the Pd is important to explain the lower energy of
the structure with two vacancies per period (RvLv) with respect to the
structure with one vacancy per period (Rv). Due to an increased number
of degrees of freedom of the RvLv structure, the Pd in that structure
is able to stretch more towards the vacancy line at the center of
figure~\ref{fig:mismatch_struct}(b), with a significant related decrease
in energy compared to the Rv structure~\cite{note_energy_decrease}. 
We stress, however, that the analysis is different when comparing the
stability of the double-vacancy-line structures (such as Rvv) and
single vacancy-line per period structures (such as Rv), which have
completely different Pd domain sizes (at constant $\eta$). An
intrinsic stability of the double-vacancy line appears to dominate the
energy trend in
that case. 

\subsection{DFT surface energies: PS {\it vs.} vacancy-line structures}

\begin{figure}[t]
\begin{center}
\includegraphics[width=7.7cm,angle=0]{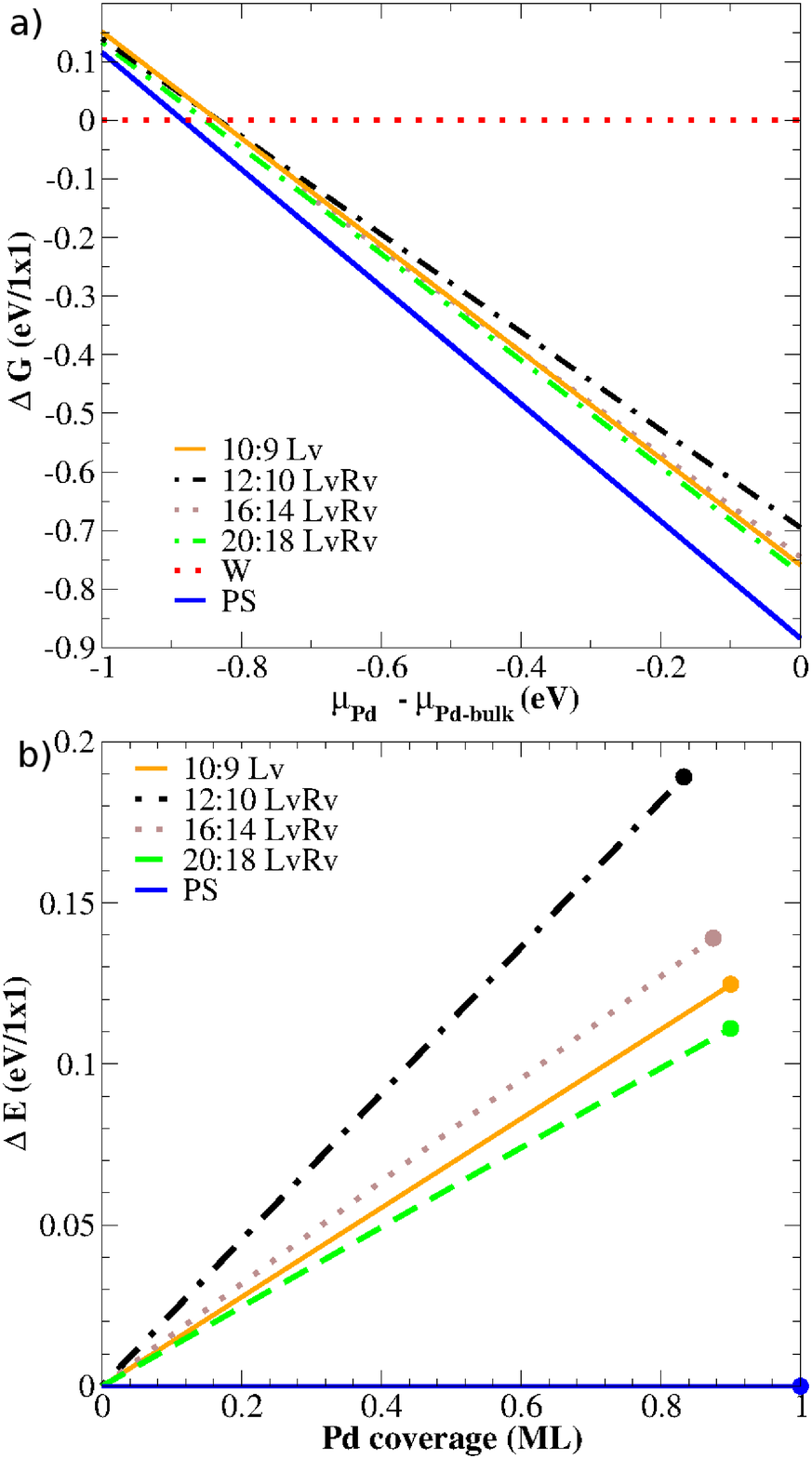}
\caption{ Calculated relative formation energies at $T=0$~K for
  various structures of Pd on W(110). (a) Formation energies of the
  single-phase surface structures  as a function of Pd chemical 
potential, (b) energy variation of the two-component Pd/W(110) and
W(110) surface as a function of Pd coverage, for different Pd/W(110)
structures, relative to the PS structure.  The circles in (b)
denote points with nominal coverage. }
\label{fig:energy}
\end{center}
\end{figure}

To determine the relative stability of the different single-phase surface
structures with different nominal coverages ($\eta$), in 
figure~\ref{fig:energy}(a) we compare the surface formation energies,
relative to the clean W(1$\times$1) surface, of the PS configuration,
two RvLv configurations 
and one Lv configuration, as a function of Pd chemical potential.

The formation energy, $\Delta G$, shown in figure~\ref{fig:energy}(a),
is   defined as~\cite{NorSchKar91}:
\begin{equation}
\Delta G =  E_{Pd/W}^{\rm (N_{\rm Pd}+N_{\rm V})x1} - (N_{\rm Pd}+N_{\rm V}) E_{\rm W}^{\rm 1x1} - N_{\rm Pd} \mu_{\rm Pd},
\end{equation}
where $E_{Pd/W}^{\rm (N_{Pd}+N_V)x1}$ stands for the energy of the Pd/W slab 
(5 layers W + 1 adsorbate Pd layer), $ E_{\rm W}^{\rm 1x1}$ for the
energy of the 5-layer W slab, $\mu_{\rm Pd}$ for the Pd chemical
potential, $N_{\rm Pd}$ and $N_V$ for the number of adsorbate atoms
and vacancy lines ($N_V=1$ or $ 2$) respectively, in the surface unit cell of the
vacancy-line structure. The superscripts denote
the sizes of the respective surface cells. In
figure~\ref{fig:energy}(a), $\Delta G$ is  normalized to the
number of W surface unit cells ($N_{\rm Pd}+N_{\rm V}$) within the (super)structure.

As seen in figure~\ref{fig:energy}(a), the PS configuration
is the ground state for all Pd chemical potentials considered. 
Vacancy  configurations are ordered in $\Delta G$ by increasing nominal
coverage $\eta$, so that, for the same type of vacancy, the structures with
longer periods are lower in energy.
This indicates that the differences in figure~\ref{fig:energy}(a) are
dominated by the energy cost of the creation of a vacancy line. The
particular type of the vacancy
line has a smaller influence on the energy ({\em e.g.}, 10:9Rv vs
20:18RvLv). The symmetric configurations are not included in the
figure, as their dependence on Pd chemical potential is identical to
that of the Rv-type (RvLv-type for the double vacancy) structures,
with a rigid shift in energy displayed in figure~\ref{fig:size_effect}. 

To address the energy changes of the surface as a function of Pd
coverage, it is useful to evaluate the internal surface energy per Pd
atom, $\Delta \varepsilon$, of the different single-phase Pd/W(110)
structures relative to the uncovered W surface:
\begin{equation}
\Delta \varepsilon\ =\ \frac{1}{N_{Pd}} \left[  E_{Pd/W}^{\rm
    (N_{Pd}+N_V)x1}\                 -\ \left(N_{Pd}+N_V\right)
                                   E_W^{1\times1}   \right].
\end{equation}
It allows one  to describe the physical situations in which only a
part of the whole W surface is covered by a uniform distribution of Pd
(in a given structure). The energy change of the whole W surface as a
function of Pd coverage may then be  evaluated as a weighted average
of the Pd-covered regions as: 
\begin{equation}
\Delta E = \Delta \varepsilon \cdot {\rm Pd\mbox{-}coverage}.
\end{equation}
In figure~\ref{fig:energy}(b), we display the $\Delta E$ surface energy
variations, as a function of Pd coverage, for the different Pd/W(110)
structures, relative to the PS structure.
The energy is normalized to the total number of W unit cells on the
surface (counting also the uncovered W).  
As  superstructures have a lower Pd density compared to the
pseudomorphic structures, they cover a larger area
(greater ``areal coverage'') for
a given submonolayer coverage of Pd, with respect to the PS
configuration. This brings about a gain in energy, 
because the uncovered tungsten surface is energetically less favorable
compared to all the Pd adlayer structures considered, as seen in
figure~\ref{fig:energy}(a). 
A superstructure with a unit cell $(N_{Pd}+N_V)\times1$ covers
$(1+N_V/N_{Pd})$ times more area compared to the pseudomorphic
layer. Therefore, in order not to include adatoms on top of the first
Pd overlayer, we limited ourselves to coverages up to the nominal
coverage $\eta$ of each structure, ({\it{i.e.}}, up to 
0.9~ML for 10:9Rv and 20:18RvLv and up to 0.83~ML for 12:10RvLv). Here, once 
again, the PS configuration has the lowest internal 
energy and the structures are ordered in $\Delta E$ by their  dislocation
density for each vacancy type. 

 The weighted-area approach used for the $\Delta E$ in
 figure~\ref{fig:energy}(b) assumes large continuous areas covered
or uncovered by the Pd adlayer,  as no effect of boundary energy (Pd
step energy on  W(110)) is taken into account. For easier comparison,
the corresponding  differences in internal energy $\Delta \varepsilon$ between the vacany-line  and
the PS structure 
are reported in table~\ref{table:Emismatch_vs_Eps}. As can be seen
from  figure~\ref{fig:energy}(b) and table~\ref{table:Emismatch_vs_Eps}, energy
differences are smaller compared to
figure~\ref{fig:energy}(a), but the structures
are ordered energetically in the same way with the PS being the lowest
energy state.

\renewcommand{\baselinestretch}{1}
\begin{table}[hb]
\bigskip
\begin{center}
\caption{Relative internal surface energies $\Delta \varepsilon$ of
  specified superstructures  with respect 
  to the PS structure. }
\begin{tabular}{ c c}
\hline
\hline \\
Superstructure & $\Delta \varepsilon$ (meV/Pd atom)\\ [2mm]
\hline  \\[-3mm]
 $20:18~RvLv$       & $24$  \\
 $16:14~RvLv$       & $32$ \\
 $10:9~Rv$            & $40$  \\
 $12:10~RvLv$       & $49$  \\
\hline
\end{tabular}
\label{table:Emismatch_vs_Eps}
\end{center}
\end{table}

Therefore, on the basis of figure~\ref{fig:energy}(a) ($\Delta G$ as a
function of Pd chemical potential) and figure~\ref{fig:energy}(b)
($\Delta E$ as a function of Pd
coverage), the ground state of Pd/W(110) at $T=0$~K is the PS
configuration, even for submonolayer coverages. This is consistent
with the observation that the experimental period increases
continuously with decreasing temperature, and tends to the PS
structure when extrapolating the trend to 0~K.  Our calculations are
also in agreement with Bauer and van der Merwe~\cite{BauMer86}, who
predicted  that at 1~ML coverage, Pd/W(110) should be
pseudomorphic. However, the
experimental findings, reported in figures~\ref{fig:exp} and
\ref{fig:exp_temp}, show that at elevated temperatures and at
submonolayer coverages the loosely-packed vacancy-line structures
prevail over the dense pseudomorphic one. Although our DFT
calculations clearly address only the internal energy (and not the
free energy at finite temperature), also these experimental findings
may be rationalized, in part, based on our DFT results.
The increase of temperature,
in general, is expected to enhance the vacancy creation, which is a
prerequisite for the vacancy-line structures. 
Furthermore,  we see from
Table~\ref{table:Emismatch_vs_Eps}, that the internal surface energy
difference between, {\it e.g.}, 20:18 RvLv and PS  is rather small
(24~meV/Pd atom), it is less than $k_BT$ ($\sim$56~meV) at 650~K, the
temperature at which a configuration with such a period and coverage
was detected. Therefore, it is reasonable to expect, based on the
energy differences, that  temperature
will influence the type of surface atomic
structure established, by making possible the formation of such
vacancy-line structures.  In addition, it can be also expected, purely
on the basis of  energy comparison among the various vacancy-line structures, that, as
temperature increases, also the structures with gradually decreasing periods can
be formed.

\subsection{Origins of the vacancy-line structures}

So far, we have not yet discussed how the vacancy-line structures
form.
A very important ingredient in deciding the structure is the
long-ranged vacancy-vacancy interactions. In reality, the corresponding
contribution to the total energy is  
inherently taken into account in our DFT slab calculations with the
relaxed atomic positions.  
Such interactions were found to be responsible for the
formation of periodic vacancy lines in similar 
adsorbate layers~\cite{CheWuZha94,ZanLouHeg95}.
In the case of  Ge/Si(001),  it was found, using STM and Monte Carlo
simulations, that  perpendicular to the vacancy lines their
interaction is short-range repulsive and long-range attractive,
balancing at $n\cdot a_0$, where $a_0$ is the Si lattice
spacing~\cite{CheWuZha94}. 
This kind of interaction has a potential
to balance the vacancy-creation energy.

However, in our case, at zero temperature, the energy gain due to vacancy-vacancy
interactions (estimated as a few tens of meV per supercell, {\it e.g.}
for the 16:14 structure, based on the relative RvLv energies in Table~\ref{table:Emismatch_vs_Eps})
is expected to be small compared to the vacancy creation energy (of at
least a few tenths of an eV per vacancy, based on the absolute
energies in Table~\ref{table:Emismatch_vs_Eps}).
The importance of temperature is revealed in the comparison of the
competing energy scales. 
Virtually all the elasticity-induced self-organized adlayers reported
in literature require elevated temperatures  
for the formation of the periodic patterns. The underlying reason is
the different scaling of the various 
contributions to the free energy with increasing thermal disorder. The
thermal scaling of the free energy parameters 
of the stripe-forming elastic lattice was given by Mente\c{s} et
al. near the critical regime for the lattice gas 
transition~\cite{MenLocAba08}. The general argument,
in which the interaction energy between the defect lines goes down
slower than the defect formation energy with increasing 
temperature, applies also to the current system with periodic
vacancy lines.

\section{Surface stress and mesoscopic stripes}

\subsection{Influence of vacancy lines on surface stress: DFT results}

\begin{figure}[t]
\begin{center}
\includegraphics[width=8cm,angle=0]{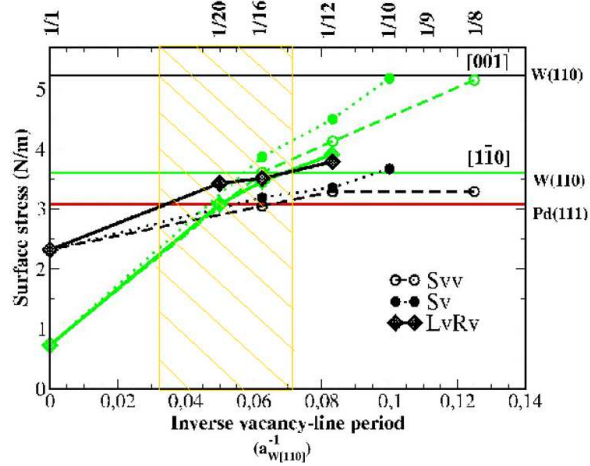}
\caption{ Calculated Pd/W(110) surface stresses for various structures along
  [1$\bar{1}$0] are shown in green (gray) and
along [001] in black, as a function of the inverse period of the
adsorbate structures in units of W(110) lattice spacings along
[1$\bar{1}$0].  The corresponding vacancy-line period n is indicated by the inverse
period fraction, 1/n, reported on the upper horizontal axis.   Lines connecting points 
are just guide to the eye. Red (dark grey) horizontal line
denotes the isotropic surface stress of the Pd(111) surface, while
stresses for the  W(110) surface along [1$\bar{1}$0] are shown in
green (light grey) and along [001] in black.  The hatched area
indicates the  interval of inverse periods which were experimentally
reached (figure~\ref{fig:exp_temp}.)   The legend is pertinent to both
crystalline directions.  }
\label{fig:stress}
\end{center}
\end{figure}

The calculated stresses for PS and various vacancy-line structures are shown in 
figure~\ref{fig:stress} as a function of the 
inverse vacancy-line period. We use here the period as a reference,
instead of coverage, as it facilitates following the trends with
supercell size for a given type 
of vacancy.  For completeness, we also include
single-vacancy-per-period structures, which have thus, for the same
inverse period, a different coverage from the other structures.

We note that in terms of the sign of the absolute surface stress, all the
structures are in tension in both
crystallographic directions. Tensile stress indicates that the stress
relaxation would cause contraction of  the interatomic distances. In
the PS structure, on the basis of the bulk inplane lattice mismatch between Pd
and W  along the W[001] and $[1\bar{1}0]$ 
 directions $\epsilon_{\rm exp}^{[001]}=14.9$~\%,
$\epsilon_{\rm exp}^{[1\bar{1}0]}=-6.2$~\%, ($\epsilon_{\rm
  theor}^{[001]}=14.4$~\% and $\epsilon_{\rm
  theor}^{[1\bar{1}0]}=-6.6$~\% in section II.B),
 one would expect a strong compressive change of stress along
[001] with respect to the stress of W(110) and a
tensile change along [1$\bar{1}$0].  Instead, from our calculations, it follows
that in both directions
the change with respect to W(110) is compressive, and of almost equal
magnitude for the PS structure. (For further discussion see
Supplementary Data).  It is a direct 
indication that the lattice mismatch cannot explain even qualitatively
the behavior of the surface stress of this monolayer adsorbate system. This result is consistent with
other studies of surface stress of monolayer adsorbates like, for example, Fe/W(110) and
Ni/W(110)~\cite{SanSchEnd98,SanEndKir99,MeySanPop03,StoBin12}.

From figure~\ref{fig:stress} we see that the stress is increasing (it becomes more
tensile) in both directions, upon addition of  vacancy lines.  Indeed,
a more tensile stress can be expected as a consequence of stretching (introduction of 
vacancy). This clearly demonstrates that the appearance of the
vacancy-line structures is not driven by  
a reduction of surface stress, a finding similar to the previous
results on the missing-row reconstruction 
at the (110) surface of fcc
transition metals~\cite{OliTreSau06}.

Importantly, we observe that the stresses for the RvLv and the Svv and Sv
structures do not differ significantly for a given period (see 1/16
and 1/12 inverse periods), especially for the cases with same Pd
coverage.  The lines connecting points have similar slopes   
and they seem more or less only off-set by a
constant value, comparable to the numerical accuracy. 

Moreover, we note that the stress changes more along  [1$\bar{1}$0], 
than along [001]. This is to be expected as the defect (and change in 
strain) is introduced 
along  [1$\bar{1}$0]. As the inverse defect period 
increases, so does the effect of the vacancy on the stress, because
the Pd stretching increases (see figure~\ref{fig:3fold_pos}). 
The change along [001] can be analysed
in terms of Poisson ratio, which states that for 
a positive Poisson ratio~\cite{positive_Poisson} stretching of an
elastic layer material along one direction would 
result in contraction of the layer along the  perpendicular  
 direction. A tendency to contract along  [001] would thus increase the 
tensile stress along that direction. 

In addition, we observe that the stress $\tau$ is less anisotropic for
vacancy-line structures with periods within the experimental range, 
than for the PS configuration. We also note that the surface tension, defined as
$(\tau_{[001]} + \tau_{[1\bar{1}0]})/2$, is significantly closer to the Pd(111) value
for inverse periods of 20, 16 and 12,
than it is for the PS configuration.  Interestingly, it appears that
the surface stress of  superstructures in the experimentally
relevant range of inverse periods is close  to the isotropic surface stress of
Pd(111) (3.08~N/m in our calculations, see in
figure~\ref{fig:stress}).

\subsection{Conditions for the appearance of linear mesoscopic
  stripes}
 
Results from figure~\ref{fig:stress}  show that the surface stresses of
the vacancy-line configurations with periods compatible with the
experimental range differ significantly from that of the PS
structure. According to the continuum model~\cite{GaoLuSuo02}, the
parameter that 
determines  the type of adsorbate stripe pattern is the ratio of the stress
changes between the two adsorbate phases along the two
crystallographic directions. The stress-change at the
boundary of the alternating regions within the stripe phase ({\em i.e.}, Pd
covered and uncovered regions on W(110)) is given by $\Delta\tau_{x} =
\tau^{Pd}_{x} - \tau^W_{x}$, where x indicates the direction, either
[$001$] or [$1\bar{1}0$]. The ratio $r = \Delta\tau_{[1\bar{1}0]} /
\Delta\tau_{[001]}$  is compared to the critical ratio $r_c = -1 +
2\nu$~\cite{GaoLuSuo02}, where $\nu$ is the W Poisson ratio (yielding $r_c=-0.44$).
 The ratio $r \approx +1$ gives stripes with no preferred orientation
(isotropic pattern of worm-like stripes), $1> r > r_c$ yields instead
linear [$1\bar{1}0$]-oriented stripes, whereas $1/r_c < r < r_c$ results in a
herringbone-like structure and  $1/r_c > r$ yields
[001]-oriented stripes. Inspection of the data presented in
figure~\ref{fig:stress} shows that the pseudomorphic Pd layer would
result in a ratio $r \approx +1$, {\em i.e.}, an isotropic pattern with worm-like
stripes, which is not observed. Instead, the vacancy-line structures
with period of 16 unit cells (which roughly corresponds to the period
observed experimentally at the temperature of
850~K), yield $r \approx 0 > r_c$, which is consistent with
the  experimentally observed mesoscopic [$1\bar{1}0$]-linear 
stripe phase. Note that, using the same formulation and the stress
  parameters from the simulated  vacancy-line
structure and for the O/W surface~\cite{StoMenBin10}, the creation of the
  experimentally-observed [$001$]-oriented  Pd-O stripes on 
W(110) has been quantitatively explained~\cite{MenStoLoc11}.

Interestingly, from figure~\ref{fig:stress} it
can be deduced that all the superstructure periods  in the
experimentally observed range do satisfy the above criterium for
[$1\bar{1}0$]-oriented Pd stripes, first
with positive values  $(r \approx 0.3)$ for the periodicity of 20 unit
cells, and then with vanishing $r$ (period of 16  unit cells),
while $r$ becomes
negative just out of this interval of experimentally
observed periods, for the periodicity of 12 unit cells. 
Upon further reduction of the period to 10 unit cells, 
$r$ becomes smaller than  $r_c$, leading thus  to a  herringbone-like
structure. 
 These predictions do not depend significantly on the type of
double vacancy structure (RvLv or Svv), while they are different for the
single-vacancy-per-period configurations.

The preceding discussion reveals that the mesoscopic stripe phases of
Pd on W(110) are sensitively dependent on the period of internal
vacancy-line structures of 
the Pd layer. The mechanism which controls the period $\lambda$  of the mesoscopic
stripes is the temperature-dependent competition between the elastic
and short-range atomic interactions~\cite{MenLocAba08}. The
characteristic length scale of the pattern is determined by $\lambda
\sim e^{(C^*_1/C_2)+1}$, where $C^*_1 = C_1(1-\frac{T}{T_c})$ is
  the boundary free energy and $T_C$ is
the Pd disordering temperature ($T_c
\approx$1200~K)~\cite{MenLocAba08}. $C_1$ is the formation energy of
the stripe boundary, {\it i.e.}, Pd step 
which in our DFT calculations comes at a cost of about $C_1 \approx
0.05$~eV/\AA, and $C_2$
is the elastic energy parameter~\cite{VanBarFei04}:
\begin{equation}
C_2 = (\Delta\tau)^2 \frac{(1-\nu^2)}{\pi {\rm E}} ,
\label{eq:elasticEnergy}
\end{equation}
with E the Young's modulus of the substrate. Taking $\nu = 0.28$ and $E = 411$~GPa for tungsten, 
and using our calculated $\Delta\tau_{[001]} = 1.7$~N/m for the 16:14
RvLv structure,
we find $C_2 = 1.3$~meV/\AA, for the mesoscopic stripes along
$[1\bar{1}0]$. Therefore, for the mesoscopic stripes 
the ratio $C_1 / C_2$ is found to be nearly 40. Thus, at $\rm T << T_c$, the characteristic length
scale of the stripe period, which is proportional to 
$e^{C_1 / C_2}$,  is astronomically large using the numbers
above. This is in agreement with the experiment, where no
pattern is observed at low temperatures.
On the other hand, for temperatures nearing the lattice gas
transition $T_c$, the step free energy, $C_1^*$, scales towards zero, and 
it becomes possible to observe the stripe phase with the period in the submicron range.

Although vacancy-line configurations might be viewed as the striped
pattern, strictly speaking, the same formulation as for the mesoscopic
stripes cannot be applied, as they do not feature a well-defined atomic step or
boundary, nor is the description within the continuum model  clearly
applicable to this system due to the atomic-scale dimensions. However,
previously, the same formalism has been applied with 
some success also to very similar systems~\cite{ZanLouHeg95,CroGirPre02}.
We have obtained a good fit (not shown) of the period {\it vs.}
temperature (in the range 400-960~K), which yielded the $C_1/C_2$ ratio of $\approx 3$.
This is an order of magnitude less than the ratio for the Pd/W(110) linear
stripes~\cite{MenLocAba08}. As the defect energy
 is comparable to the formation energy
of the Pd step, the long-range term, regarding the vacancy-vacancy
interactions, should be an order of magnitude stronger.
This indicates that the period of the vacancy-line structures
should be about an 
order of magnitude  smaller than the typical period of the
mesoscopic stripes and at temperatures much lower than the lattice gas
disordering transition, which is
fully consistent with the experimental observations.
 A more rigorous proof would require a discretized version of the continuum model of
elastic interactions~\cite{AleVanMea88} with an arbitrary force
distribution within the period structure.

Finally,  we conclude this section by noting that the vacancy-line order along
[1${\bar1}$0] is a prerequisite for the formation of linear mesoscopic
stripes. In particular, in our calculations, only the vacancy-structure
periods of 12 unit cells and somewhat larger 
yield the right ratio of stress differences in the two
directions. This is in agreement with the experimental findings which
for Pd/W(110) yield a period of 14 unit cells along $[1\bar{1}0]$ for
the internal structure of Pd mesoscopic
linear stripes, while in the case of the Pd-O stripes on W(110) the Pd
vacancy-line periods are longer~\cite{MenStoLoc11}.

\section{Conclusion}

We studied  the interplay between the surface structure and  stress in
Pd/W(110) for submonolayer coverages of Pd. In particular, we have
characterized experimentally and theoretically the Pd/W(110) 
superstructure which, at high temperature (above 900~K), is observed as the
 internal structure of the Pd
mesoscopic stripes. We  demonstrated  its critical role on the formation of
the observed linear mesoscopic stripes.

The structure of Pd/W(110) was investigated using LEED and LEEM,
 revealing a particular temperature dependence of the
periodic superstructure inverse period in the measured range (400 to
$\sim$1200~K). In order to analyze further the observed
superstructures, we performed DFT simulations of the
internal-structure configurations.
The calculations considered the PS structure along with periodic
vacancy-line structures expected in a loosely-packed layer  along
$[1\bar{1}0]$. Among all the configurations theoretically considered,
the PS one was singled out as the lowest energy state at 0~K, which is
consistent with the LEED inverse-period trend extrapolated to low
temperature. The
experimental observation of the periodic superstructure instead of the
PS layer at elevated temperatures was explained qualitatively to be due
to vacancy formation and vacancy-vacancy interactions becoming more
pronounced at increased temperatures.
Finally, the calculated surface stresses for the
vacancy-line structures were found to significantly differ from the stress of
the PS configuration. Among different vacancy-line
configurations with the same period in the experimental range, instead,
the surface stress remained comparable. On the basis of
the continuum model theory and our surface-stress values, we
demonstrated that the vacancy-line structure in a certain range
of periods is a prerequisite for the formation of the linear mesoscopic
stripes, thus resulting in a peculiar arrangement of two mutually
perpendicular types of striped periodic orderings at different lenght
scales.

\ack
We acknowledge E.~Bauer for
providing the thesis~\cite{Geo_thesis} and we thank him and 
A.~Locatelli for useful discussions.  
The calculations were mostly performed on the sp6 supercomputer at CINECA, under a
grant ``Computational studies of self-assembling and induced magnetism
at metal surfaces''.

\section*{References}



\bibliography{inco}

\begin{thebibliography}{10}
\bibitem{note_transition}
The onset of the transition towards the pseudomorphic structure is determined
  by fitting a constant function, in the range of smaller coverages, and the
  function $\rm \alpha/(1-coverage)$, for larger coverages, to the period vs
  coverage curve in figure~\ref{fig:exp}. The constant, parameter $\alpha$ and
  the kink point at which the two ranges of the functions meet are used as free
  parameters. The latter is taken as the transition coverage.
\bibitem{note_d_H_site}
The interlayer distances are similar for the H site: $d_{12}=-5.9$~\% and
  $d_{23}=-0.5$~\%.
\bibitem{note_degrees_freedom}
We note that in our systems the degrees of freedom of the particles to be
  considered for the energy normalization when comparing to the thermal energy
  are those of the Pd atoms (as opposed to those of the vacancy/defect). In
  fact, all Pd atoms in the unit cell relax differently and participate in the
  structural rearrangement of the Pd overlayer upon creation of the vacancy
  lines (see end of Section IV B).
\bibitem{note_energy_decrease}
An estimated energy decrease by 11~meV/(1$\times$1) is obtained by moving the
  Pd atoms adjacent to the central vacancy line in figure~\ref{fig:3fold_pos}
  from the $d_x$ distance found in the Rv 6:5 (Lv 6:5) structure (with
  relaxed $d_z$ position) to the $d_x$ distance in the RvLv 12:10 structure,
  with all other atomic positions frozen in the relaxed RvLv 12:10
  configuration. The estimate of 11~meV, obtained by considering only the
  effect of displacement of the Pd atoms adjacent to the central vacancy line,
  is about half of the total energy difference between the Rv and RvLv
  structures (23 meV/(1$\times1$)).
\bibitem{positive_Poisson}
Although the exact Poisson ratio is unknown for this surface, we can speculate
  that it is most likely indeed positive based on the fact that it has a
  regular atomic structure (most of the systems with negative/anomalous Poisson
  ratio have complicated structures due to their uniquely oriented molecular
  orbitals) and on its behavior perpendicular to the surface: we note that the
  contraction of the Pd adlayer perpendicular to the slab is largest for small
  periodicities (largest strain) and is decreasing towards larger periodicities
  and the PS structure, as shown in Fig.~S3(a), and as expected for positive
  Poisson ratio.
\end{thebibliography}

\end{document}


\beginsupplement

\title[Supplementary Information for ``Self-organization of Pd/W(110)...'']{Supplementary Information for ``Self-organization in Pd/W(110): 
interplay between surface structure and stress''}

\author{N.~Stoji\'{c}$^{1,2}$, T.~O.~Mente\c{s}$^3$ and N.~Binggeli$^{1,2}$}

\address{$^1$Abdus Salam International Centre for Theoretical Physics, 
Strada Costiera 11, Trieste 34151, Italy}
\address{$^2$ IOM-CNR Democritos,  Trieste, I-34151, Italy}
\address{$^3$ Sincrotrone Trieste S.C.p.A., Basovizza-Trieste 34012, Italy}

\section{ Diffraction spot width}
 
\begin{figure}[ht]
\begin{center}
\includegraphics[width=11cm, angle=0]{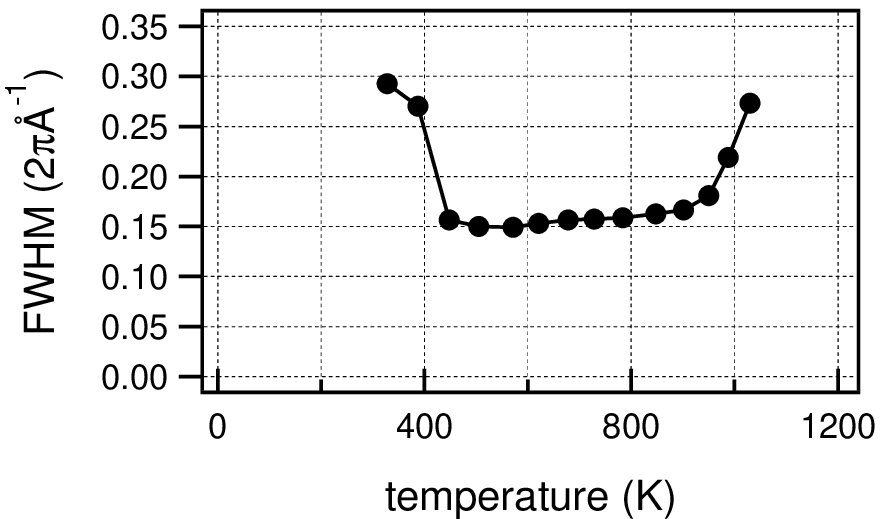}
\caption{Diffraction spot width of the Pd superstructure as a function
of temperature at 0.5~ML coverage.}
\label{fig:fwhm}
\end{center}
\end{figure}

In order to facilitate analysis of Fig.~2 from the main text,
describing the inverse period as a function of temperature, we
have also analyzed the diffraction spot width (full width at half
maximum, FWHM) from LEED, as shown
in Figure~\ref{fig:fwhm}. 

In the range between 400 and 900~K, the superstructure spot has an
almost constant width. This changes dramatically upon decrease of
temperature below 400~K, where the full width at half maximum
increases sharply. Similarly, an increase in FWHM is observed
above 900~K.

\section{ Surface stress for the hollow-site PS structure}

Initially, when starting the DFT study of Pd/W(110), we had assumed
the hollow site was the ground state for the Pd PS overlayer on
W(110). The convergence of the surface stress with slab thickness was
investigated thus using this reference structure. 

\begin{figure}[ht]
\begin{center}
\includegraphics[width=7cm, angle=270]{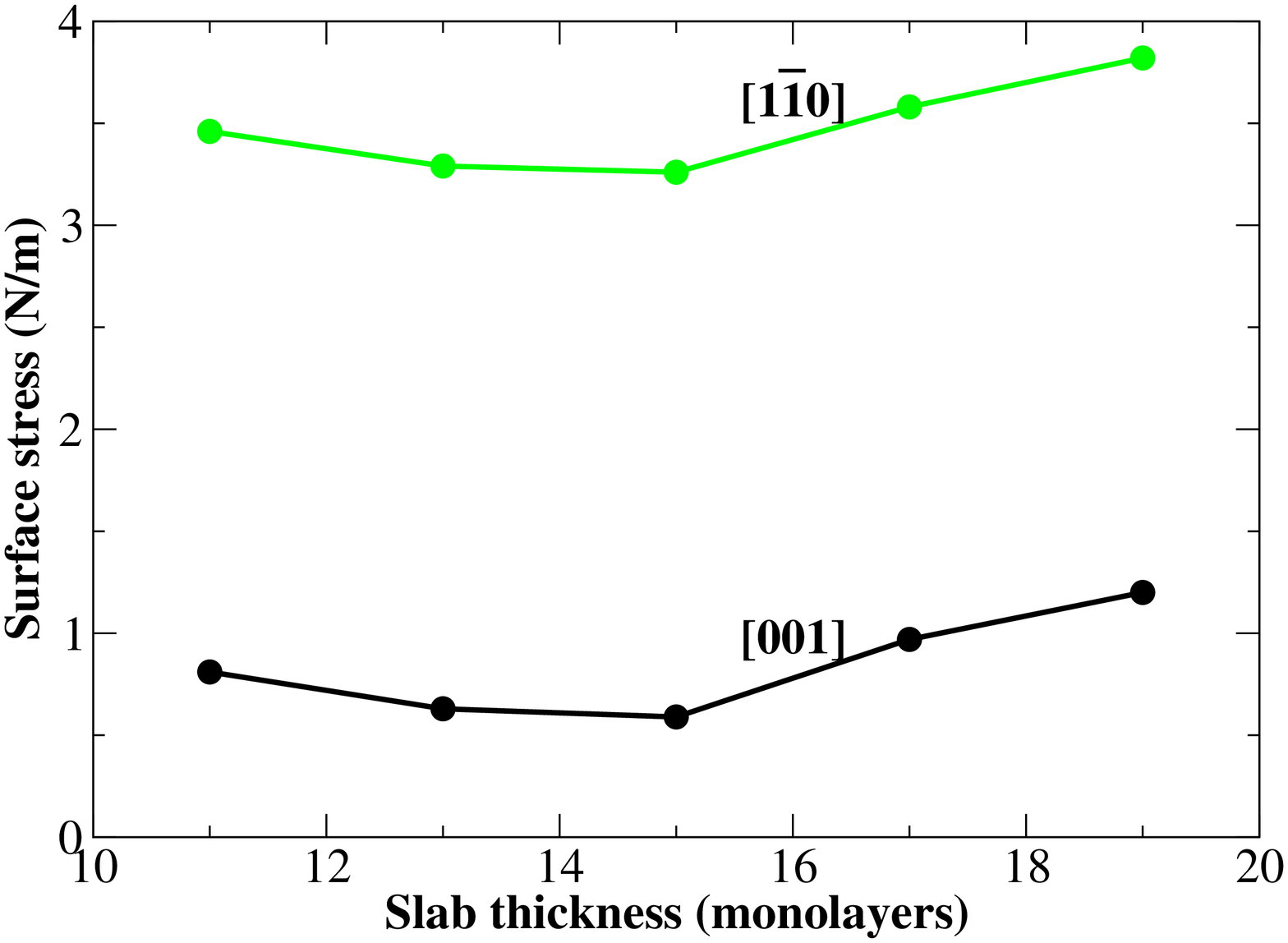}
\caption{ The Pd/W(110)/Pd surface-stress convergence with  slab
  thickness for the Pd PS configuration with the H adsorption site.   }
\label{fig:slab_thickness}
\end{center}
\end{figure}

The surface stress dependence on  slab thickness is shown 
in figure~\ref{fig:slab_thickness} for symmetric  Pd/W(110)/Pd
slabs. The changes of surface stress among the slabs between 11 and 19
layers 
can be as large as 0.6~N/m.  Even so, the stress anisotropy, which we define as
 $\tau_{[1\bar{1}0]}-\tau_{[001]}$ is constant within less than
 0.1~N/m, which gives us confidence  
that, as long as we perform all the calculations at the same
thickness, we can compare the stresses 
of different structures with significant precision. 

Interestingly, the surface stress of the PS structure critically
depends on the adsorption site. 
When comparing the stress components for the hollow and  L (R) adsorption site, the values
for 11 layers from figure~\ref{fig:slab_thickness} should be compared with  
figure~9 from the main text. It can be seen right
away that there are not only large quantitative differences between
the stresses of two PS
configurations, but also a reversal in the stress anisotropy
($\tau_{[1\bar{1}0]} > \tau_{[001]}$ for the hollow site, whereas
$\tau_{[001]} > \tau_{[1\bar{1}0]} $ for the 3-fold site). 
 For the hollow-site PS, the stress along  [1$\bar{1}$0] is
more than 3 times larger (or 2.7~N/m), than for the 3-fold site, while
along  [001] the stress for the 3-fold site is about 1.5~N/m larger
than for the hollow site. This is not so
surprising given  the Pd position in the 3-fold adsorption site. 
Along  [1$\bar{1}$0], the Pd atom is very close to one W atom, which
produces a large compressive stress.  Along  [001] both bond angle and
length change, resulting in a more tensile stress for the 3-fold
site.

\section{ Atomic relaxations of the various vacancy-line structures}

\begin{figure}[t*]
\begin{center}
\includegraphics[width=7.5cm, angle=0]{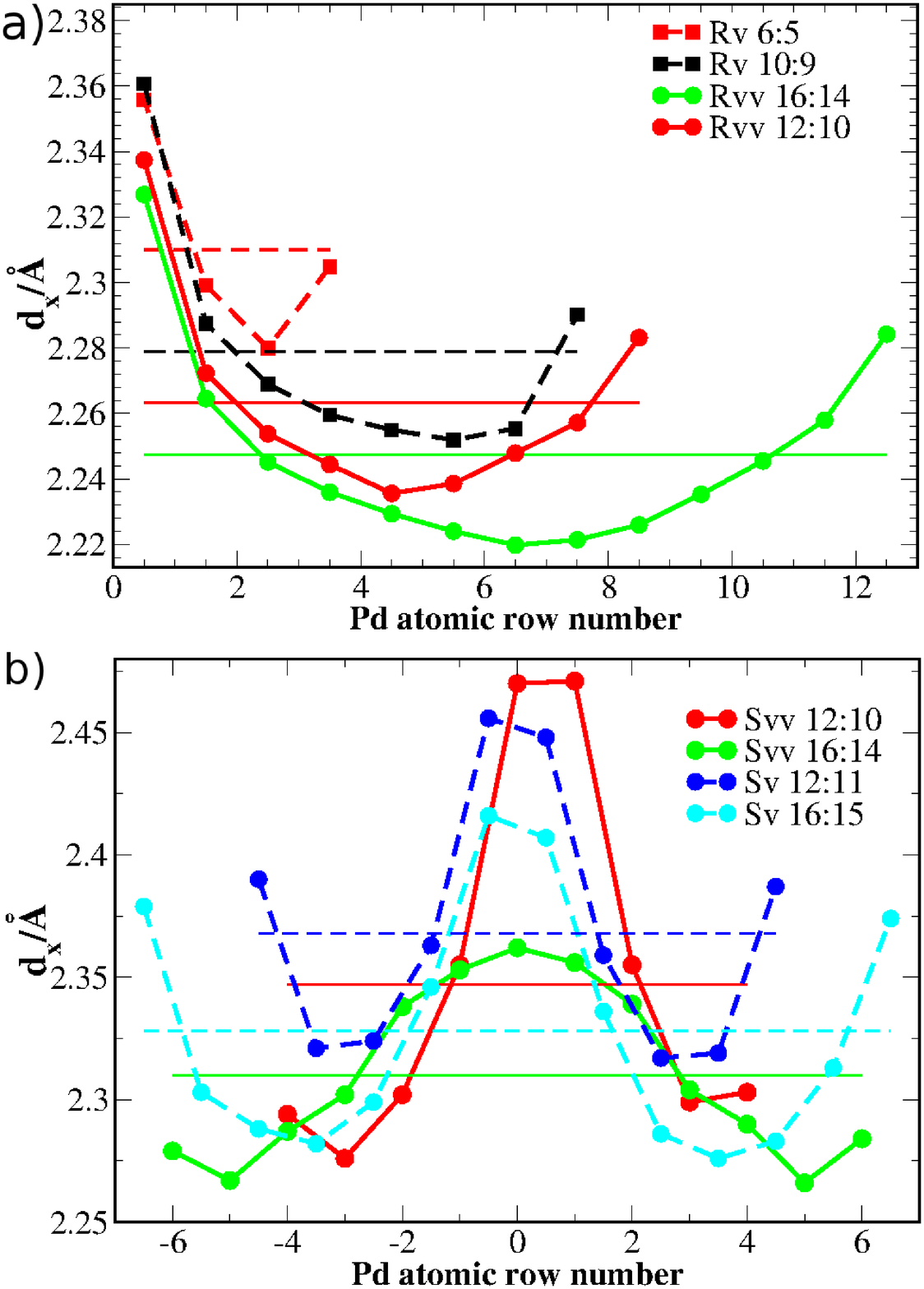}
\caption{Interatomic spacing, $d_x$, along  [1$\bar{1}$0] between rows
  of Pd atoms for the R-type structures (Rv and Rvv) (a) and for the
  S-type structures (Sv and Svv) (b).  Straight lines indicate  the
average values.}
\label{fig:pos}
\end{center}
\end{figure}

In figure~\ref{fig:pos} we present atomic relaxations along
[1$\bar{1}$0], in terms of interatomic distances, $d_x$, analogous to
figure~7(a) from the main text, for the remaining structures of the R-type
(Rv and Rvv) in (a), and of the S-type (Sv and Svv) in (b). 
``Atomic row number'' describes the atomic row position within the
Pd structure along [1$\bar{1}$0]. For the
description of the Rv and Rvv structures
(figure~\ref{fig:pos}(a)), the zeroth atomic row number was set at the
first row of Pd atoms,  just after the vacancy. 
For all the Rv and Rvv structures described, the edges of the Pd
structure relax the most, but there is also  a significant asymmetry
between the zeroth and the last atom in-plane relaxations,
{\it i.e.}, the zeroth atom has larger displacement. The relaxation of
the last Pd atom is limited by the vicinity, in the [1$\bar{1}$0]
direction, of the W atom in the plane below. Overall, between the Rv and
RvLv type of structures, the relaxations
along [1$\bar{1}$0] are always larger on average in the RvLv case for the same
coverage $\eta$, see Sec. IVB. The Rvv structures are characterized
with yet smaller in-plane relaxations, due to a larger Pd domain size
within the structure for the same coverage (for the coverage of {\it
  e.g.}, $\eta=0.833$, the Rvv domain has width of 10 atomic rows along
[1$\bar{1}$0], while Rv and RvLv have 5 atomic-row wide domains). 
For the symmetric structures, in figure~\ref{fig:pos}(b), the zeroth atomic
row is placed on the central Pd atom within the surface structure in
the case of the odd number of Pd atoms (Sv structures) and in-between
the two central Pd
atoms for the even number of Pd atoms (Svv structures). The in-plane
relaxations are largest for the atoms in the center of the structure
(most likely due to the unfavorable adsorption site),
while also the edge atoms have significant displacement along
[1$\bar{1}$0], comparable to the relaxations of the
R-type structures at the vacancy between the R and L-site Pd domains. 
The least stretched are the atoms between the central
and edge parts, sitting near 3-fold positions. 
In general, the in-plane relaxations for the Sv and Svv structures are
considerably larger, than for the
RvLv, Rv and Rvv structures for the same coverage.

\begin{figure}[t]
\begin{center}
\includegraphics[width=7cm, angle=0]{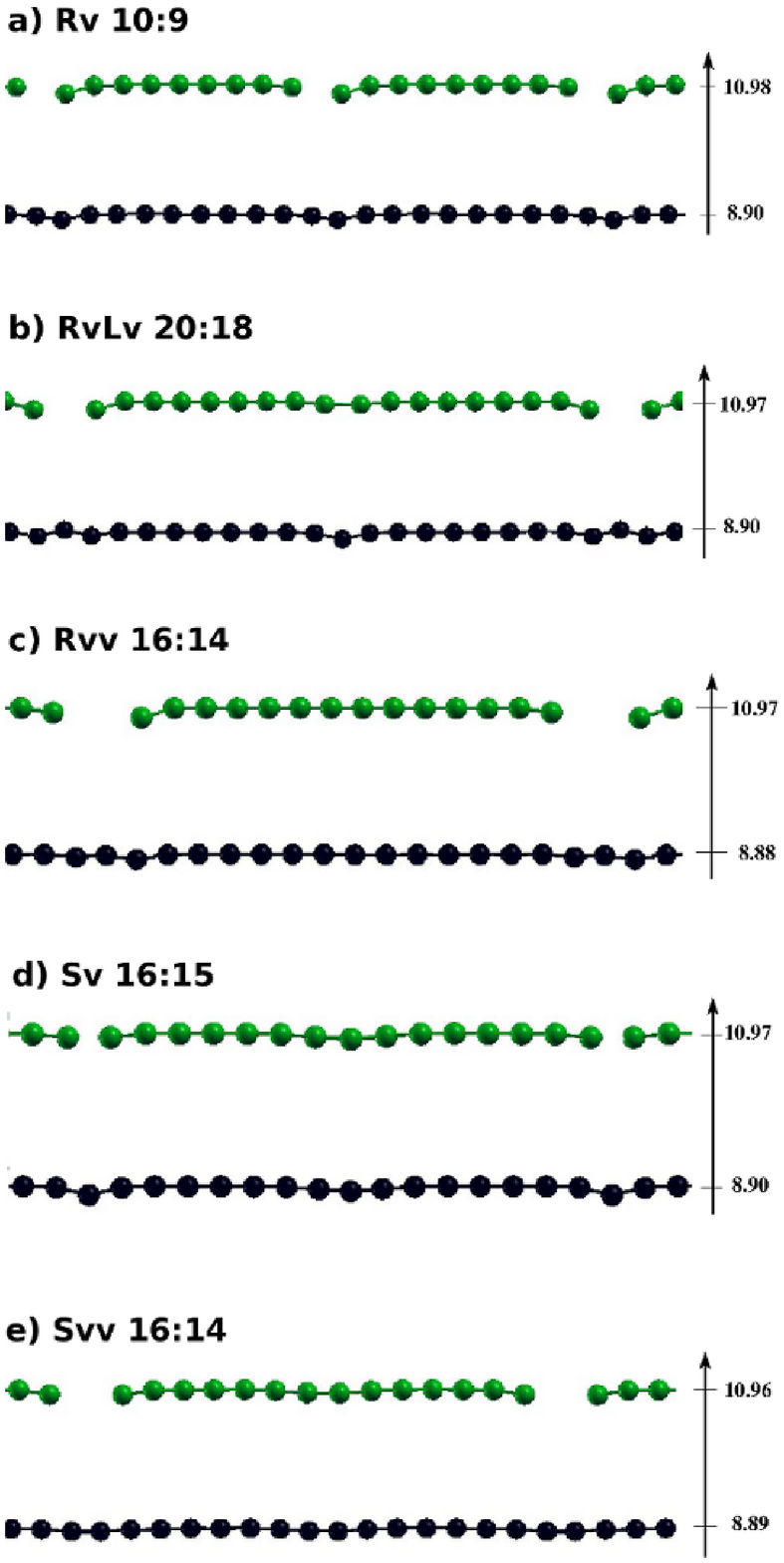}
\caption{Atomic positions along the perpendicular direction for the
  upper two layers (Pd + W) for  different vacancy-line structures.
  Pd and W atoms are shown in green (grey) and black. The  
  positions of the outermost Pd and W atom in \AA~ with respect to the
  bottom of our calculational slab are indicated at the arrows. The
  reference values for the PS  structure are 10.95 and 8.86~\AA.}  
\label{fig:profile}
\end{center}
\end{figure}

Figure~\ref{fig:profile} shows the atomic positions along the
perpendicular direction for the Pd and the underlying W layer, for
the different vacancy-line structures from figure~5 in the main text.
The Rv, RvLv and Rvv structures are characterized by strong perpendicular
displacement at the edges of the Pd domains (about 0.13~\AA~ in all
three structures). We note that for the
RvLv  structures, the perpendicular Pd displacement at the second vacancy (between
the Pd domains with the R and the L sites) is much smaller than at the
first vacancy (between the L and the R-site-domains).  The downward
motion at the second vacancy is diminished by the lateral proximity of the W
atom underneath. Similarly, in
the Rv case, the perpendicular Pd displacement is asymmetric, as the
edge Pd atoms on two sides of the Pd domain have different
environment; on one side, the edge Pd atom is close to the H site and
can relax more than the edge Pd atom on the other side, where the underlying W atom is
very close. 
The Sv structure
shows the same trend of relaxation at the edges, but of a more moderate size, while the Svv one
has the Pd edge 
displacements comparable ($\approx$0.07~\AA) to the structures of the R-type.  

 Interestingly,
also the underlying W atoms displace significantly in the
perpendicular direction, mostly at the edges of the Pd
domains for all the structures in Figure~\ref{fig:profile} (less so
for the Svv structure). The size of the W displacement in this layer,
close to the vacancy is about 0.08~\AA.
 The W motion is present even in the case when
there is no significant perpendicular displacement of the Pd atom, as
for the second vacancy of the RvLv structure. In that case, the Pd
atoms have a very strong in-plane relaxation, almost covering the
underlying W atom, which then moves inward. 
 This W displacement underneath the Pd-domain edges is inward for all the structures (of about
 0.07-0.10~\AA, except for the Sv structure, for which it is smaller), while for the RvLv and Rv
structures, in addition, the uncovered W atom moves considerably
outwards (up to 0.10~\AA). The outward motion can be rationalized in
terms of the
in-plane motion of the neighboring W atoms, which both follow the
motion of the Pd atom above, squeezing the W atom in-between, which
then moves upwards.  

 We note that also the second W layer (the third from the surface overall)
relaxes in the perpendicular direction. For example, the outmost W atom in this layer is 6.67~\AA~
away from the bottom of the slab in the Rvv 16:15 structure, with almost 0.05~\AA~ variation in
the z-direction. 
 In all the structures, the total perpendicular inward relaxation (as
 obtained from the total slab height)  is
 smaller, than for the PS configuration. Within the same type of structure,
the total inward relaxation is increasing with the period of the
structure.


\bibliography{inco}
